\documentclass[fleqn,10pt]{wlscirep}
\usepackage[utf8]{inputenc}
\usepackage[T1]{fontenc}
\usepackage{chemformula}
\let\ce\ch
\usepackage{mathtools}
\usepackage{physics}
\usepackage{floatrow}
\usepackage{subcaption}
\usepackage{hyperref}
\usepackage[justification=centering]{caption}
\usepackage{amsmath}
\usepackage{nccmath}
\usepackage{subcaption}
\usepackage{floatrow}
\usepackage[thinlines]{easytable}
\usepackage{cleveref}

\title{Radical pairs can explain magnetic field and lithium effects on the circadian clock}

\author[1,2,3,*]{Hadi Zadeh-Haghighi}
\author[1,2,3,*]{Christoph Simon}

\affil[1]{Department of Physics and Astronomy, University of Calgary, Calgary, AB, T2N 1N4, Canada}
\affil[2]{Institute for Quantum Science and Technology, University of Calgary, Calgary, AB, T2N 1N4, Canada}
\affil[3]{Hotchkiss Brain Institute, University of Calgary, Calgary, AB, T2N 1N4, Canada}

\affil
[*
]{hadi.zadehhaghighi@ucalgary.ca, csimo@ucalgary.ca}

\begin{abstract}
\textit{Drosophila}’s circadian clock can be perturbed by magnetic fields, as well as by lithium administration. Cryptochromes are critical for the circadian clock. Further, the radical pairs in cryptochrome also can explain magnetoreception in animals. Based on a simple radical pair mechanism model of the animal magnetic compass, we show that both magnetic fields and lithium can influence the spin dynamics of the naturally occurring radical pairs and hence modulate the circadian clock’s rhythms. Using a simple chemical oscillator model for the circadian clock, we show that the spin dynamics influence a rate in the chemical oscillator model, which translates into a change in the circadian period. Our model can reproduce the results of two independent experiments, magnetic fields and lithium effects on the circadian clock. Our model predicts that stronger magnetic fields would shorten the clock’s period. We also predict that lithium influences the clock in an isotope-dependent manner. Furthermore, our model also predicts that magnetic fields and hyperfine interactions modulate oxidative stress. The findings of this work suggest that quantum nature and entanglement of radical pairs might play roles in the brain, as another piece of evidence in addition to recent results on xenon anesthesia and lithium effects on hyperactivity.

\end{abstract}

\begin{document}

\flushbottom
\maketitle
\thispagestyle{empty}

\section*{Introduction}
All organisms, including microbes, plants, and animals, use an endogenous timekeeping system, namely the circadian clock (CC), which helps organisms to adapt to the 24-h cycle of the earth to control their daily physiology and behavior rhythms. Molecular pacemakers inside organisms drive the CC. In mammals, the coordination of essential behavioral, hormonal, and other physiological rhythms throughout the body relies on the CC \cite{takahashi2017transcriptional}. It is also known that the circadian clock modulates cognitive activities \cite{kyriacou2010circadian,aten2018mir,teodori2019shedding,silver2014circadian} and is linked to mood disorders \cite{mcclung2007circadian}. In \textit{Drosophila}, the CC controls the timing of eclosion and courtship, the period of rest and activity, and the timing of feeding; it also influences temperature preference \cite{allada2010circadian,zhang2012molecular}. Despite the differences in molecular components of the CCs, their features, organization, and the molecular mechanism that generate rhythmicity are very alike across organisms \cite{tataroglu2014studying}.\par

Environmental cues such as light, food, and temperature can modulate the rhythmicity of the CC \cite{patke2020molecular}. It is also known that the CC is susceptible to external magnetic fields (MFs). In the 1960s, Brown et al. \cite{brown1960response} found that small changes in the intensity of Earth's MF synchronize the CCs of fiddler crabs and other organisms. Since then, the effects of external MF on the CC have been observed in multiple studies \cite{bliss1976circadian,contalbrigo2009effects,marley2014cryptochrome,close2014compass,close2014compassb,fedele2014genetic,lewczuk2014influence,vanderstraeten2015could,manzella2015circadian,bartos2019weak,lai2021genetic,thoni2021therapeutic,xue2021biological}. Similarly, Yoshii et al. \cite{yoshii2009cryptochrome} have shown the effects of static MFs on the CC of \textit{Drosophila} and found that exposure to these fields exhibited enhanced slowing of clock rhythms in the presence of blue light, with a maximal alteration at 300 $\mu$T, and reduced effects at both lower and slightly higher     field strengths. However, the exact mechanism behind this phenomenon is still mostly unknown. \par

Additionally, a growing body of evidence points to the circadian cycles as a target for bipolar disorder treatments \cite{Abreu2015,Takahashi2008}. Bipolar disorder is correlated with disruptions in circadian rhythms \cite{Abreu2015,Takahashi2008} and abnormalities in oxidative stress \cite{Yumru2009,Salim2014,MachadoVieira2007,Ng2008,Andreazza2008,Lee2013,Brown2014,Berk2011}. Lithium is the first-line treatment for bipolar disorders \cite{Yin2006,Li2012}, yet the exact mechanisms and pathways underlying this treatment are under debate. It has been shown that lithium treatment for hyperactivity in rats is isotope dependent \cite{Ettenberg2020}. Lithium has two stable isotopes, \ce{^{6}Li} and \ce{^{7}Li}, which have different nuclear spin angular momentum, $I_6=1$ and $I_7=3/2$, respectively. In a recent study, it has also been proposed that lithium affects hyperactivity via the clock center channel in the brain \cite{zadeh2021entangled}. This study further predicted that the magnetic field would influence the potency of lithium treatment. Furthermore, Dokucu et al. \cite{dokucu2005lithium} showed that in \textit{Drosophila} lithium lengthens the period of the CC. Here, based on these findings, we propose a mechanism that can explain both the MF effects and lithium effects on \textit{Drosophila}'s CC.\par

\begin{figure}
     \begin{subfigure}{0.5\linewidth}
        \centering
        \includegraphics[width=0.9\textwidth]{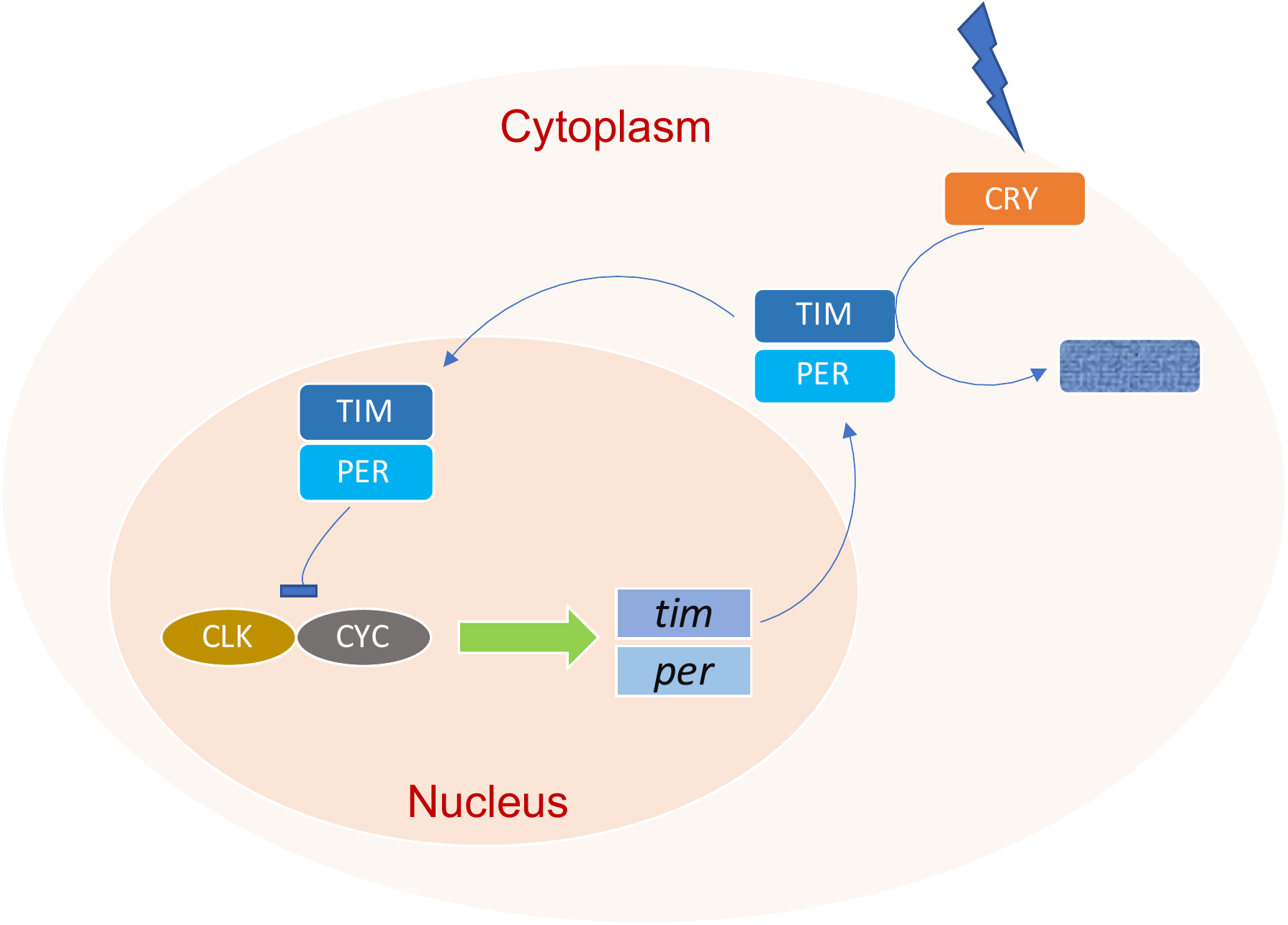}
        \caption{}
        \label{fig:cc-schem}
    \end{subfigure}
    \hfill
\begin{minipage}{0.4\linewidth}
    \begin{subfigure}{\linewidth}
\hspace*{-1cm}
\includegraphics[width=1.2\textwidth]{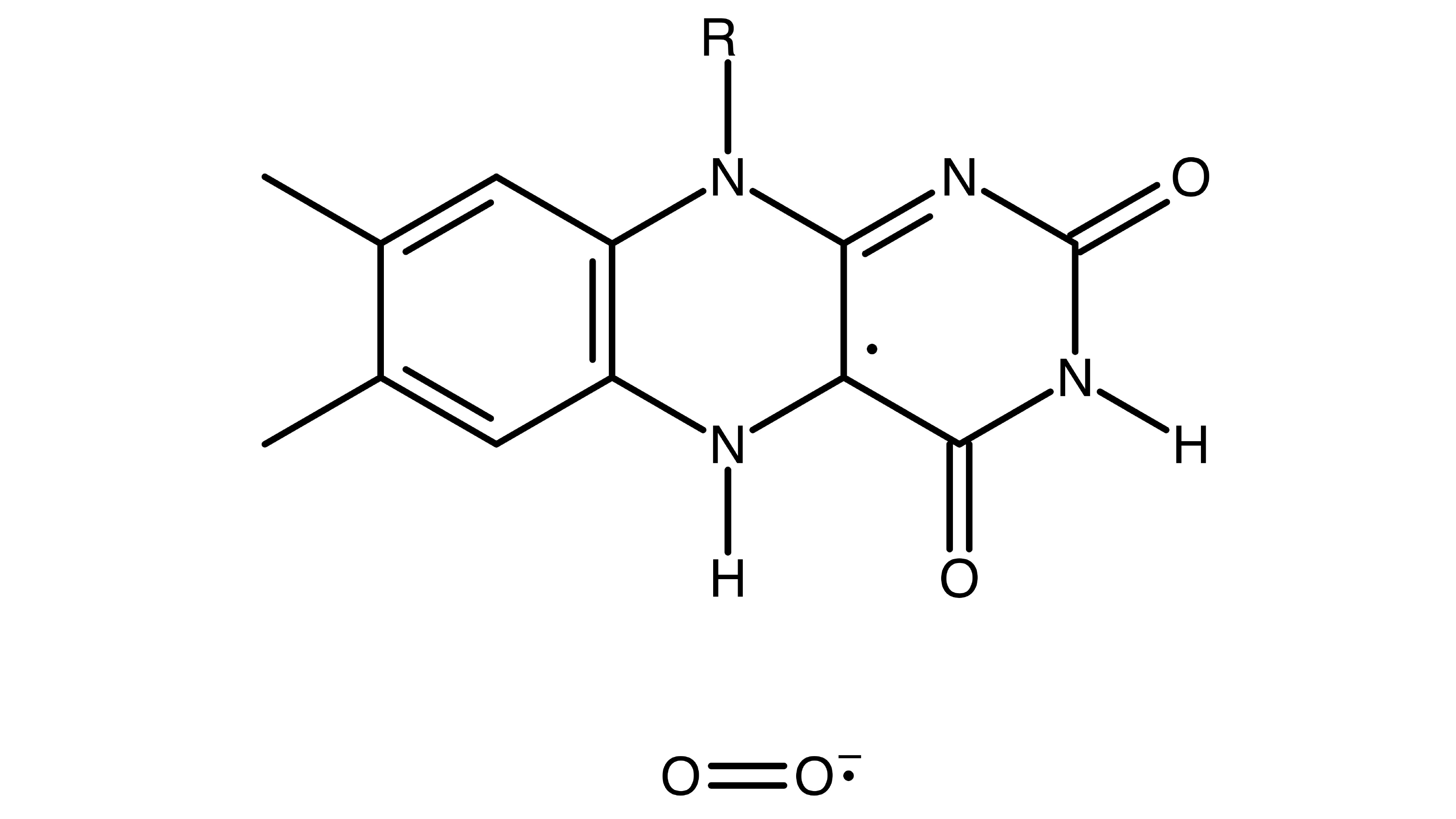}
        \caption{}
        \label{fig:fadh-o2}
    \end{subfigure}
\end{minipage}%
\caption{a) Simplified models of the circadian clock feedback loop in \textit{Drosophila}. CLOCK (CLK) and CYCLE (CYC) promote the $tim$ and $per$ genes. PER and TIM
first accumulate in the cytoplasm and then enter into the
nucleus to block their gene transcription. Upon light, absorption CRY binds to TIM, and this results in the degradation of TIM \cite{tyson1999simple,leloup1999limit}. b) Flavinsemiquinone, \ch{FADH^{.}}, and superoxide \ch{O2^{.-}} radical pair in CRY, considered in the RPM model in the present work. The radical pair undergoes interconversion between singlet and triplet states.}
\label{fig:schem}
\end{figure}

In the CC of \textit{Drosophila}, the CLOCK (CLK) and CYCLE (CYC) transcription factors form a heterodimeric complex and promote the period ($per$) and timeless ($tim$) transcription mRNAs, which result in the assembly of the PERIOD (PER) and TIMELESS (TIM) proteins in the cytoplasm \cite{tataroglu2015molecular}, shown in Fig. \ref{fig:cc-schem}. During the night, PER and TIM accumulate and form a heterodimer. The TIM/PER complex enters the nucleus and then promotes the phosphorylation of CLK/CYC, which inhibits the promotion of the $per$ and $tim$ mRNAs. During the day, TIM and PER are gradually degraded, and consequently, CLK/CYC are released from repression to start a new cycle. \par 

In this process, light activation of cryptochrome (CRY) protein is critical for the rhythmicity of the CC. CRYs regulate growth and development in plants; they also act as photo-receptors in some animal's CC, where they are necessary components of the circadian clock \cite{emery2000drosophila,chaves2011cryptochromes}. In \textit{Drosophila}'s CC, upon light absorption, CRY undergoes a conformational change that allows it to bind TIM \cite{ceriani1999light,allada2010circadian,patke2020molecular} which results in the degradation of TIM and hence resetting the clock, see Fig. \ref{fig:cc-schem}. CRYs contain the flavin adenine dinucleotide (FAD) cofactor, which is the photoreception segment. Upon blue-light absorption, FAD can go through various redox states. In insects, this process produces the anionic semiquinone FAD, \ch{FAD^{.-}}, and reactive oxygen species (ROS), which are thought to be the key signaling states for initiating TIM degradation \cite{ozturk2011reaction,vaidya2013flavin}. In mammals, CRY's are essential for the development of intercellular networks in the suprachiasmatic nucleus (SCN), a circadian pacemaker in the brain, that subserve coherent rhythm expression; the network synchronizes cellular oscillators and corrects errors \cite{Welsh2010}.\par 

It has been known for many years that migratory birds use Earth’s magnetic field for finding their way during migrations. Later, it was proposed that the radical pair mechanism (RPM) could be the key for birds’ magnetoreception \cite{hore2016radical}. Ritz et al. \cite{ritz2000model} proposed that the candidate protein for such a mechanism could be the CRY in the retina of birds. Ever since, there have been extensive studies on that hypothesis, and to date, it is the most promising model for avian magnetoreception \cite{xu2021magnetic,wan2021cryptochrome,Jones2016} in birds, sharks, sea turtles, monarch butterflies, fruit flies, etc. These models are based on the study of the dynamics of the created pair of radicals which can be in a superposition of singlet (S) and triplet (T) states \cite{Steiner1989}, depending on the parent molecule's spin configuration \cite{Timmel1998}. The key elements in such reactions are radical molecules—transient molecules with an odd number of electrons in their outer molecular shell. Protons, neutrons, and electrons possess spin angular momentum, an inherently quantum characteristic. In a simple picture, quantum spins are like tiny magnets; any other spins or magnetic field in the vicinity could alter their states. In the framework of RPM for avian magnetoreception, it is thought that in CRY RPs can be in the form of anionic semiquinone FAD radical (\ch{FAD^{.-}}) and terminal tryptophan radical (\ch{TrpH^{.+}}) \cite{Giovani2003,hore2016radical,Hong2020,hochstoeger2020biophysical}. It is also well-known that the superoxide radical, \ch{O2^{.-}}, can be an alternative partner for the flavin radical \cite{Mller2011,Romero2018,Chaiyen2012,Mondal2019}. It has also been proposed that RPs can play important roles in the magnetosensitivity of \textit{Drosophila}'s CC \cite{bartos2019weak,yoshii2009cryptochrome,gegear2010animal}.\par 

Applied magnetic fields can also influence oxidative stress in the presence of CRY \cite{sherrard2018low}. Moreover, the CC rhythmicity is associated with an endogenous rhythm in the generation of ROS \cite{edgar2012peroxiredoxins}. Furthermore, redox signaling rhythms are intrinsically coupled to the circadian system in mammals \cite{pei2019diurnal}. Thus it seems pertinent to explore the connection between magnetic field effects and ROS role in the CC. It has recently been proposed that radical pairs (RPs) could play roles in other brain functions. Dufor et al. \cite{Dufor2019} propose that weak MFs activate cellular signaling cascade in neural circuits via acting through CRY, most likely by modulating the state of RPs. The authors concluded that the presence of CRY is critical in axon outgrowth under low-intensity repetitive transcranial magnetic stimulation (rTMS). It has also been suggested that RP may help explain xenon-induced anesthesia \cite{Smith2021} and the lithium effects on mania \cite{zadeh2021entangled}. It, therefore, seems RPs could play critical roles in the functionalities of the brain in general and the CC in particular.\par

The circadian oscillations in \textit{Drosophila} can be modeled by incorporating the formation of a complex between the PER and TIM proteins and introducing negative feedback loops \cite{goldbeter2002computational}, which are the key to the rhythmicity of PER and TIM and their mRNA transcription. The models can be described by a set of a few kinetic equations \cite{leloup1998model}. However, modeling \textit{Drosophila} CC \cite{leloup1998model} can be further simplified into two nonlinear equations \cite{tyson1999simple}. Furthermore, Player et al. \cite{player2021amplification} show that quantum effects such as magnetic field effects and hyperfine interaction of radical pairs can be introduced to the chemical oscillator by considering the quantum effects on the corresponding reaction rates.\par

Here, we propose that the RPM could be the underlying mechanism behind the lithium treatment effects and MF effects on \textit{Drosophila}'s CC. MF via the Zeeman interaction and lithium nucleus via HFIs modulate the recombination dynamics of singlet-triplet interconversion in the naturally occurring RPs in the [\ch{FADH^{.}} ... \ch{O2^{.-}}] complex, shown in Fig. \ref{fig:fadh-o2}, and hence influence the period of the CC. \par

In the following, we review the quantitative experimental results for the effects of applied magnetic field \cite{yoshii2009cryptochrome} and lithium \cite{dokucu2005lithium} on the period of \textit{Drosophila}'s CC. Next, we briefly describe the quantum spin dynamics for the radical pair model where the magnetic field effects and the HFIs are relevant. Moving on, we present our singlet yield calculation for the RP system, inspired by the CRY-based model of birds' avian magnetoreception \cite{Hore2019}. Later we use a simple model for the mathematical presentation of \textit{Drosophila}'s CC, following the work on Tyson et al. \cite{tyson1999simple}. Then we introduce the quantum effect to the period of the CC model, and we show the consistency of our model's predictions and the experimental findings on the magnetic field and lithium treatment effects. Finally, we discuss new predictions for experiments.

\section*{Results}
\subsection*{Magnetic field and lithium treatment effects on circadian clock and RPM}
\subsubsection*{Results from prior experiments}
Here, we focus on the effects of static MF on \textit{Drosophila}'s CC observed by Yoshii et al. \cite{yoshii2009cryptochrome}. The authors conducted experiments to observe the effects of static magnetic fields with different intensities, [0, 150, 300, 500] $\mu$T, on changes in the period of \textit{Drosophila}'s CC under blue light illumination, shown in Table \ref{tab:yoshii}. These magnetic fields are, excepting the control of 0 $\mu$T, approximately 3, 6, and 10 times stronger than natural magnetic fields, respectively. That observation revealed that the period alterations significantly depended on the strength of the magnetic field such that the period change reached a maximum of 0.522$\pm$0.072 h at 300 $\mu$T. In this experiment, the geomagnetic field was shielded, and the arrhythmic flies were excluded from the analysis. We also consider the results of the experiment conducted by Dokucu et al. \cite{dokucu2005lithium} observing the effects of chronic lithium administration on \textit{Drosophila}'s CC for a range of doses [0, 300] mM. It was shown that lithium treatment lengthens the CC  with a maximum prolongation of 0.7$\pm$0.217 h at 30 mM of lithium compared to zero lithium intake, see Table \ref{tab:dokucu}. In that work, the lethality of lithium up to 30 mM was relatively low until the end of the experiments. Here, we consider that 30 mM is the optimal concentration of lithium where all RPs interact with lithium atoms. We assume that the lithium administered in that work was in its natural abundance, 92.5\% and 7.5\%  of \ce{^7{Li}} and \ce{^6{Li}}, respectively. Here we will refer to the natural lithium as \ce{Li}. In our model here, 0 $\mu$T of MF and 0 mM of lithium are our control sets for MF and lithium effects on the CC.  

\begin{table}[tbhp]
\caption{\label{tab:yoshii}Period Changes in the free-running rhythm of \textit{Drosophila} after application of magnetic fields (MFs) under blue light illumination and lithium administration, taken from the work of Yoshii et al. \cite{yoshii2009cryptochrome}.} 
\begin{tabular}{|c|c|c|}
\hline
\textbf{Applied MF $\mu$T}               & \textbf{Period Change (h)} &\textbf{Number of flies} \\ \hline
0 & 0.302$\pm$  0.052          &             27             \\ \hline
150 &            0.394              $\pm$    0.048         &              26            \\ \hline
300 &            0.522               $\pm$     0.072        &             23             \\ \hline
500) &             0.329             $\pm$      0.057        &              25            \\ \hline
\end{tabular}
\end{table}

\begin{table}[tbhp]
\caption{\label{tab:dokucu}Period in the free-running rhythm of \textit{Drosophila} for zero and 30 mM intake of lithium, taken from the work of Dokucu et al. \cite{dokucu2005lithium}} 
\begin{tabular}{|c|c|c|}
\hline
\textbf{Lithium dose (mM)}               & \textbf{Period (h)}   & \textbf{Number of flies} \\ \hline
0               &             23.7            $\pm$   0.033        &             311             \\ \hline
30               &             24.4              $\pm$    0.214         &             44             \\ \hline
\end{tabular}
\end{table}

\subsubsection*{RPM model}
We develop an RP model to reproduce static MFs and lithium administration effects on the rhythmicity of \textit{Drosophila}'s CC observed in Ref. \cite{yoshii2009cryptochrome} and Ref. \cite{dokucu2005lithium}, respectively. Taking into account the facts that the CC is associated with oxidative stress levels under light exposure \cite{ozturk2011reaction,vaidya2013flavin,emery1998cry,arthaut2017blue,edgar2012peroxiredoxins} and applied MF \cite{ehnert2017extremely,pei2019diurnal}, and the CC is affected by lithium intake, we propose that the applied magnetic field interacts with the spins of RPs on FADH and superoxide, and the nuclear spin of lithium modulates the spin state of the radical on superoxide. The correlated spins of RP are assumed to be in the [\ch{FADH^{.}} ... \ch{O2^{.-}}] form, where the unpaired electron on each molecule couples to the nuclear spins in the corresponding molecule,  see Fig. \ref{fig:fadh-o2}. In [\ch{FADH^{.}} ... \ch{O2^{.-}}]. \par

We consider a simplified system in which the unpaired electron is coupled to the flavin's nitrogen nucleus with an isotropic HF coupling constant (HFCCs) of 431.3 $\mu$T \cite{lee2014alternative}. In this model, for simplicity, we consider only Zeeman and HF interactions \cite{Efimova2008,hore2016radical}. Following the work of Hore \cite{Hore2019}, the anisotropic components of the hyperfine interactions are excluded, which are only relevant when the radicals are aligned and immobilized \cite{Schulten1976}. The RPs are assumed to have the $g$-values of a free electron. The Hamiltonian for the RP system reads as follows:    
\begin{ceqn}
\begin{equation}
    \hat{H}=\omega \hat{S}_{A_{z}}+a_{A} \mathbf{\hat{S}}_A.\mathbf{\hat{I}}_{A}+\omega \hat{S}_{B_{z}}+a_{B} \mathbf{\hat{S}}_B.\mathbf{\hat{I}}_{B},
\end{equation}
\end{ceqn}
where $\mathbf{\hat{S}}_A$ and $\mathbf{\hat{S}}_B$ are the spin operators of radical electron A and B, respectively, $\mathbf{\hat{I}}_A$ is the nuclear spin operator of the isoalloxazine nitrogen of  \ch{FADH^{.}}, similar to Refs. \cite{Hore2019,zadeh2021entangled}, $\mathbf{\hat{I}}_B$ is the nuclear spin operator of the Li nucleus, $a_{A}$ and $a_{B}$ are HFCCs, taken from \cite{lee2014alternative,zadeh2021entangled}, and $\omega$ is the Larmor precession frequency of the electrons due to the Zeeman effect. Of note, oxygen has a zero nuclear spin and thus its HFCC equals zero, ($a_{B}=0$), however in the model for lithium effects $a_{B}$ corresponds to the nuclear spin of lithium. We assumed that the RPs start off from singlet states (see the Discussion section).

\subsubsection*{Singlet yield calculation}
The singlet yield resulting from the radical pair mechanism can be obtained by solving the Liouville-von Neumann equation for the spin state of the radical pair throughout the reaction. Using the eigenvalues and eigenvectors of the Hamiltonian, the ultimate singlet yield, $\Phi_S$, for periods much greater than the RP lifetime \cite{Hore2019} has the following form:

\begin{ceqn}
\begin{equation}
    \Phi_S=\frac{1}{4}-\frac{k}{4(k+r)}+\frac{1}{M}\sum_{m=1}^{4M}\sum_{n=1}^{4M}|\bra{m}\hat{P}^S \ket{n}|^2 \frac{ k(k+r)}{(k+r)^2+(\omega_m-\omega_n)^2},
\end{equation}
\end{ceqn}
where $M = M_A M_B$, $M_{X} =\prod\limits_{i}^{N_X} I_{iX}(I_{iX}+1)$, is the nuclear spin multiplicity, $\hat{P}^S$ is the singlet projection operator, $\ket{m}$ and $\ket{n}$ are eigenstates of $\hat{H}$ with corresponding eigenenergies of $\omega_m$ and $\omega_n$, respectively, $k$ is the RP reaction rate, and $r$ is the RP spin-coherence lifetime rate (relaxation rate).\par

\begin{figure}
  \includegraphics[width=0.6\linewidth]{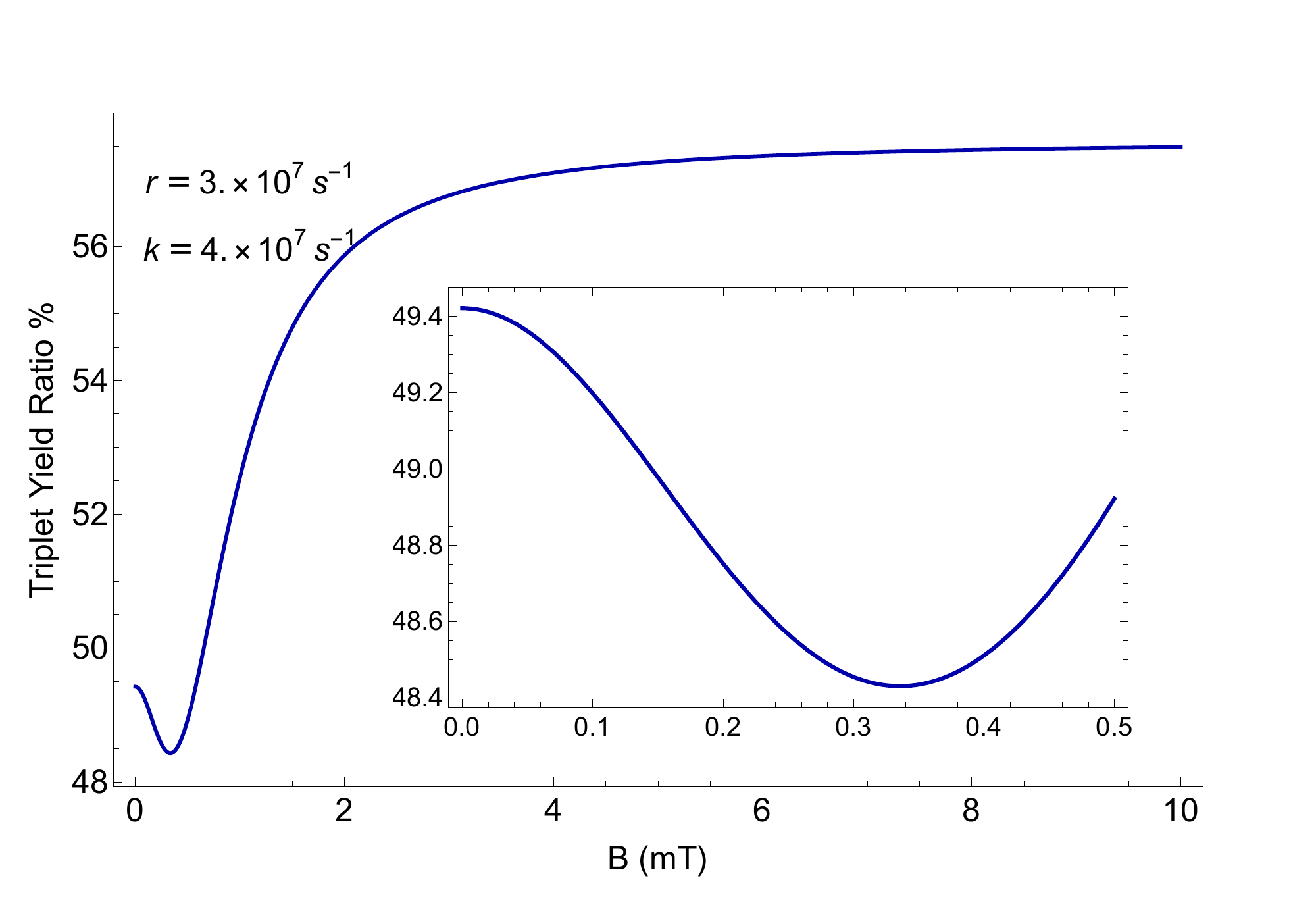}
  \caption{The dependence of the singlet yield of the [\ch{FADH^{.}} ... \ch{O2^{.-}}] complex on external magnetic field $B$ for $a_{1A}=431.3$ $\mu$T, reaction rate $k$, and relaxation rate $r$. The singlet yield reaches a minimum value of 48.45\% in [280-360] $\mu$T (see the inset).} 
  \label{fig:ty-b}
\end{figure}

Here we look at the sensitivity of the singlet yield to changes in the strength of the external magnetic field for the [\ch{FADH^{.}} ... \ch{O2^{.-}}] radical complex. Fig. \ref{fig:ty-b} illustrates the dependence of the singlet yield of the [\ch{FADH^{.}} ... \ch{O2^{.-}}] complex on external magnetic field $B$ with a maximum yield in [280-360] $\mu$T for $k=4\times 10^{7}$ s$^{-1}$ and $r=3\times 10^{7}$ s$^{-1}$ with $a_{1A}=431.3$ $\mu$T. In our model, the magnetic dependence of singlet yield is the foundation of the magnetic sensitivity of the circadian clock. Using the singlet yield, we can reproduce the experimental finding on the effects of applied MF \cite{yoshii2009cryptochrome} and lithium administration \cite{dokucu2005lithium} on the period of the circadian clock of \textit{Drosophila}, as we discuss below. It is worth mentioning that the singlet-product of the RP system in [\ch{FADH^{.}} ... \ch{O2^{.-}}] is \ch{H2O2} \cite{usselman2016quantum}, which is the major ROS in redox regulation of biological activities and signaling \cite{sies2020reactive}.

\subsubsection*{Circadian clock model}
We use a simple mathematical model for the circadian clock of 
\textit{Drosophila}, following the work of Tyson et al.\cite{tyson1999simple}. In this model, PER monomers are rapidly phosphorylated and degraded, whereas PER/TIM dimers are less susceptible to proteolysis, shown in Fig. \ref{fig:cc-schem}. In this context, it is also assumed that the cytoplasmic and nuclear pools of dimeric protein are in rapid equilibrium. With these considerations, it is possible to write the mathematical model in two coupled equations as follows:
\begin{ceqn}
\begin{equation}
\label{eq:m}
    \frac{dM(t)}{dt}=\frac{v_m}{1+(P_t(t)(1-q(t))/2P_{crit})^2}-k_m M(t),
\end{equation}
\end{ceqn}

\begin{figure}
     \begin{subfigure}{0.45\linewidth}
        \centering
        \includegraphics[width=1.\textwidth]{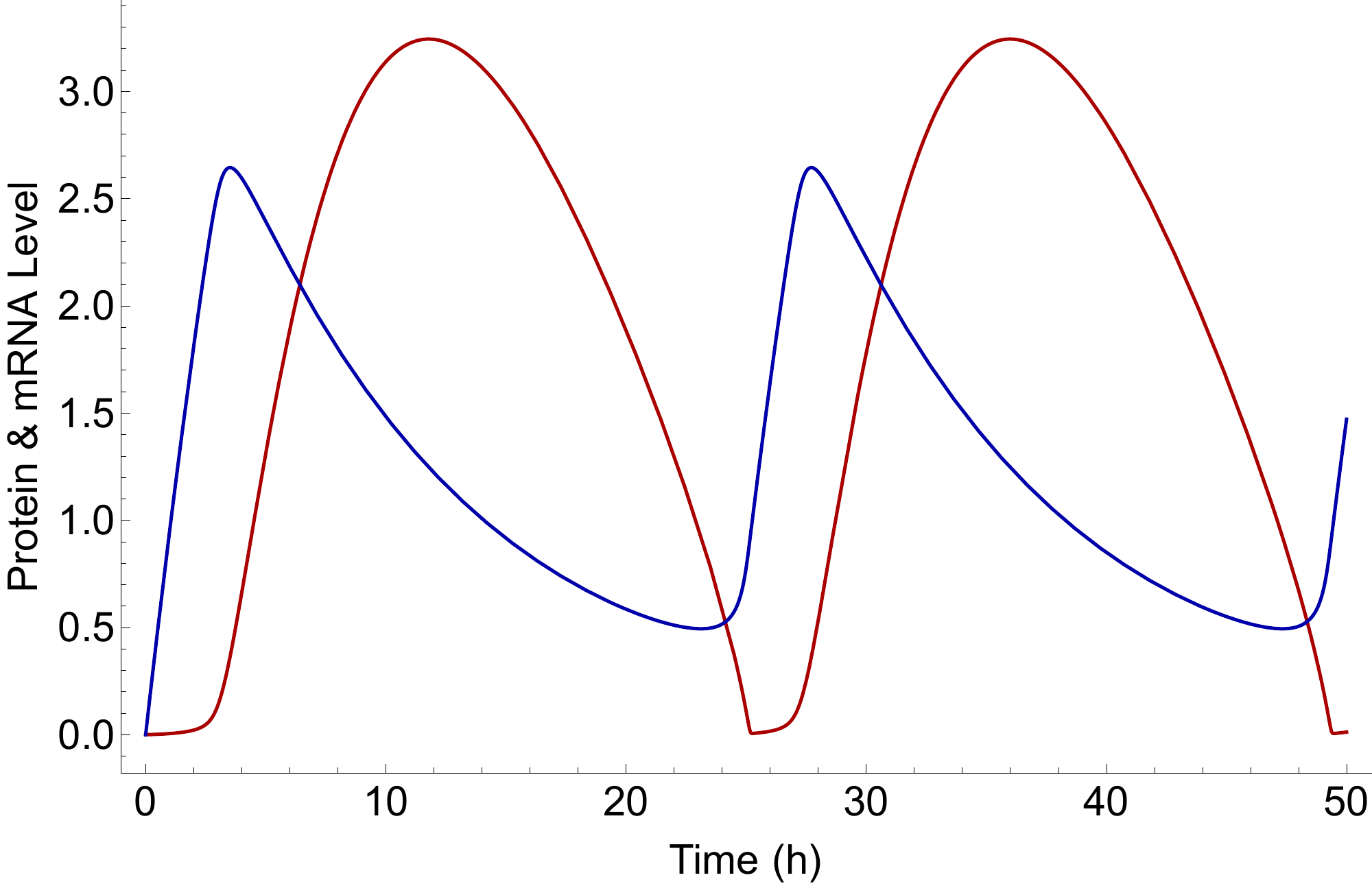}
        \caption{}
        \label{fig:pro/mrna-t}
    \end{subfigure}
    \hfill
\begin{minipage}{0.45\linewidth}
    \begin{subfigure}{\linewidth}
\includegraphics[width=1.\textwidth]{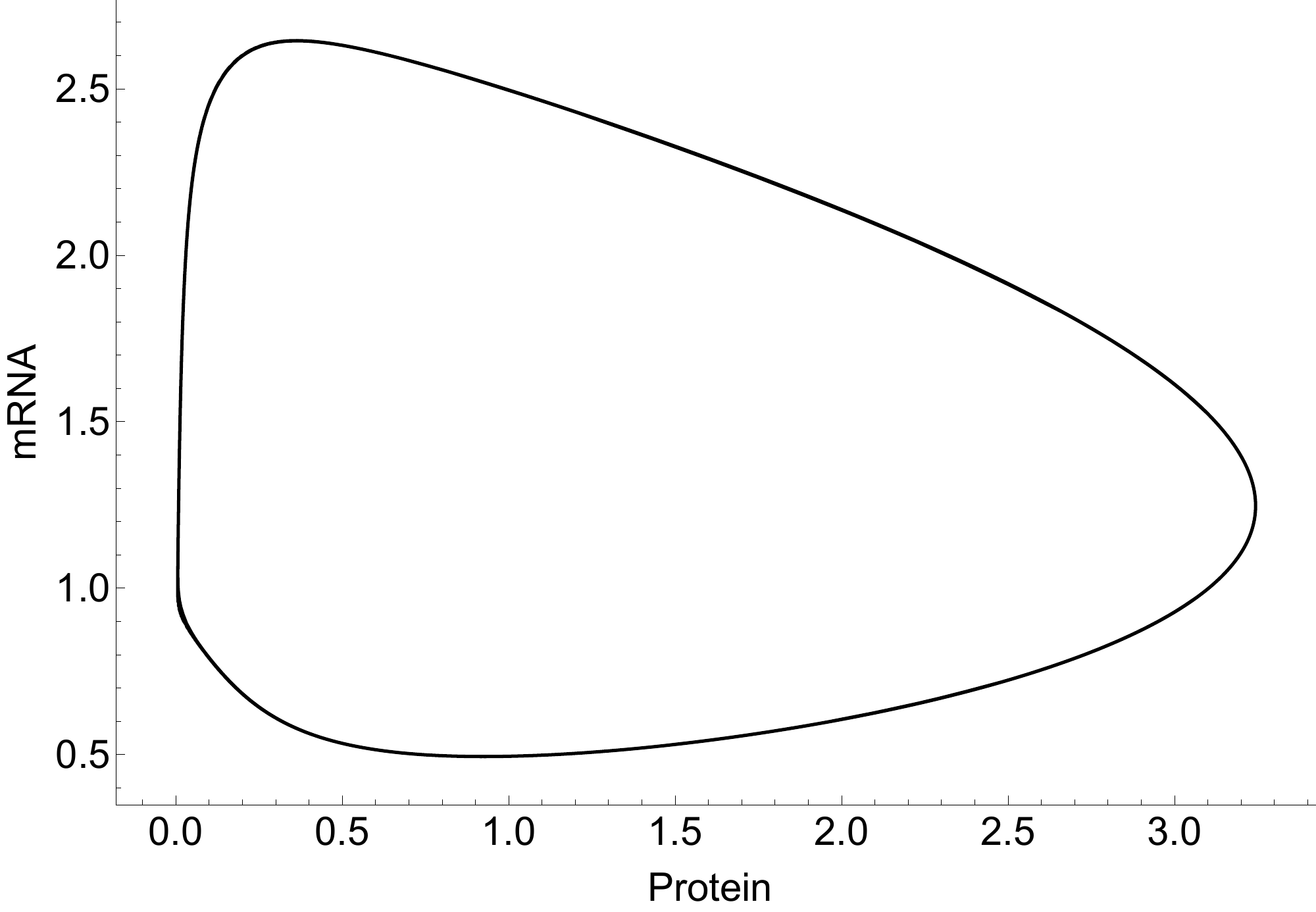}
        \caption{}
        \label{fig:pro-mrna}
    \end{subfigure}
\end{minipage}%
\caption{a) Explicit time-dependence of the concentrations of protein [red] and mRNA [blue] and b) Parametric representations of oscillations in the concentrations of protein and mRNA, shown as a limit cycle in \textit{Drosophila}'s circadian clock model using Eqs. \ref{eq:m} and \ref{eq:pt}, and the parameters from Table \ref{tab:tyson}.}
\label{fig:math}
\end{figure}

\begin{ceqn}
\begin{equation}
\label{eq:pt}
    \frac{dP_t(t)}{dt}=v_d M(t)-\frac{k_{p1} P_t(t) q(t)+k_{p2} P_t(t)}{J_p+P_t(t)}-k_{p3} P_t(t),
\end{equation}
\end{ceqn}
where $q(t)=\frac{2}{1+\sqrt{1+8K_{eq}P_t(t)}}$, $P_t(t)$ and $M(t)$ are the total protein and the mRNA concentrations, respectively. For the descriptions and values of the parameters, see Table \ref{tab:tyson}. In this simple model, $k_{p3}$ represents the role of CRY's light activation and hence proteolysis of protein. By solving Eqs. \ref{eq:m} and \ref{eq:pt}, we obtain the oscillation of protein and mRNA concentrations. Fig. \ref{fig:math} shows the explicit time-dependence of protein and mRNA concentrations and the parametric representation of the chemical oscillator limit cycle for \textit{Drosophila}’s CC. To obtain the period of the clock, we take the average differences between successive peaks and likewise troughs of either $P_t(t)$ or $M(t)$ by keeping track of when the derivative is zero. 

\begin{table}[H]
\centering
\caption{\label{tab:tyson}Parameter values for the circadian clock of \textit{Drosophila}, taken from the work of Tyson et al. \cite{tyson1999simple}. C$_m$ and C$_p$ are characteristic concentrations for mRNA and protein, respectively.} 
\begin{tabular}{llll}
\textbf{Name} & \textbf{Value} & \textbf{Units} & \textbf{Description} \\
\hline
v$_m$ &1.0 &C$_m$ h$^{-1}$  & maximum rate of mRNA synthesis \\
k$_m$ &0.1& h$^{-1}$  &mRNA degradation rate constant\\
v$_p$& 0.5& C$_p$ C$_m^{-1}$ h$^{-1}$  &mRNA rate constant\\
k$_{p1}$& 10& C$_p$h$^{-1}$  & $V_{max}$ of monomer phosphorylation\\
k$_{p2}$ &0.03 &C$_p$ h$^{-1}$  & $V_{max}$ of dimer phosphorylation\\
k$_{p3}$& 0.1 &h$^{-1}$  &proteolysis rate constant caused by CRY activation\\
K$_{eq}$ &200& C$_p^{-1}$& dimerization equilibrium constant\\
P$_{crit}$& 0.1& C$_p$ & Dimer concentration at the half-maximum transcription rate\\
J$_p$ &0.05& C$_p$ & Michaelis constant for protein kinase (DBT)\\
\end{tabular}
\end{table}

\subsubsection*{Effects of singlet yield change on circadian clock}
The effects of applied magnetic fields and hyperfine interactions can be introduced to the chemical oscillator of the circadian clock by modifying the rate $k_f$ \cite{player2021amplification}, following the work of Player et al., see Methods. In the CC Eqs. \ref{eq:m} and \ref{eq:pt} the corresponding rate is $k_{p3}$, which is 0.1 h$^{-1}$ for the natural cycle of the clock. Hence for the occasions with no singlet yields effects, this value must be retained. The singlet yield effects on $k_{p3}$ can be written as follows:

\begin{ceqn}
\begin{equation}\label{eq:yeild-chem3}
k'_{p3} \propto k_{p3} \frac{\Phi_S'}{\Phi_S},
\end{equation}
\end{ceqn}
where $k'_{p3}$, $\Phi_S$, and $\Phi'_S$ are the modified rate constant $k_{p3}$, the singlet yield with no quantum effects, and the singlet yield resulted from quantum effects due to the Zeeman and/or hyperfine interactions, respectively. \par 

\begin{figure}
    \centering
    \includegraphics[width=.7\linewidth]{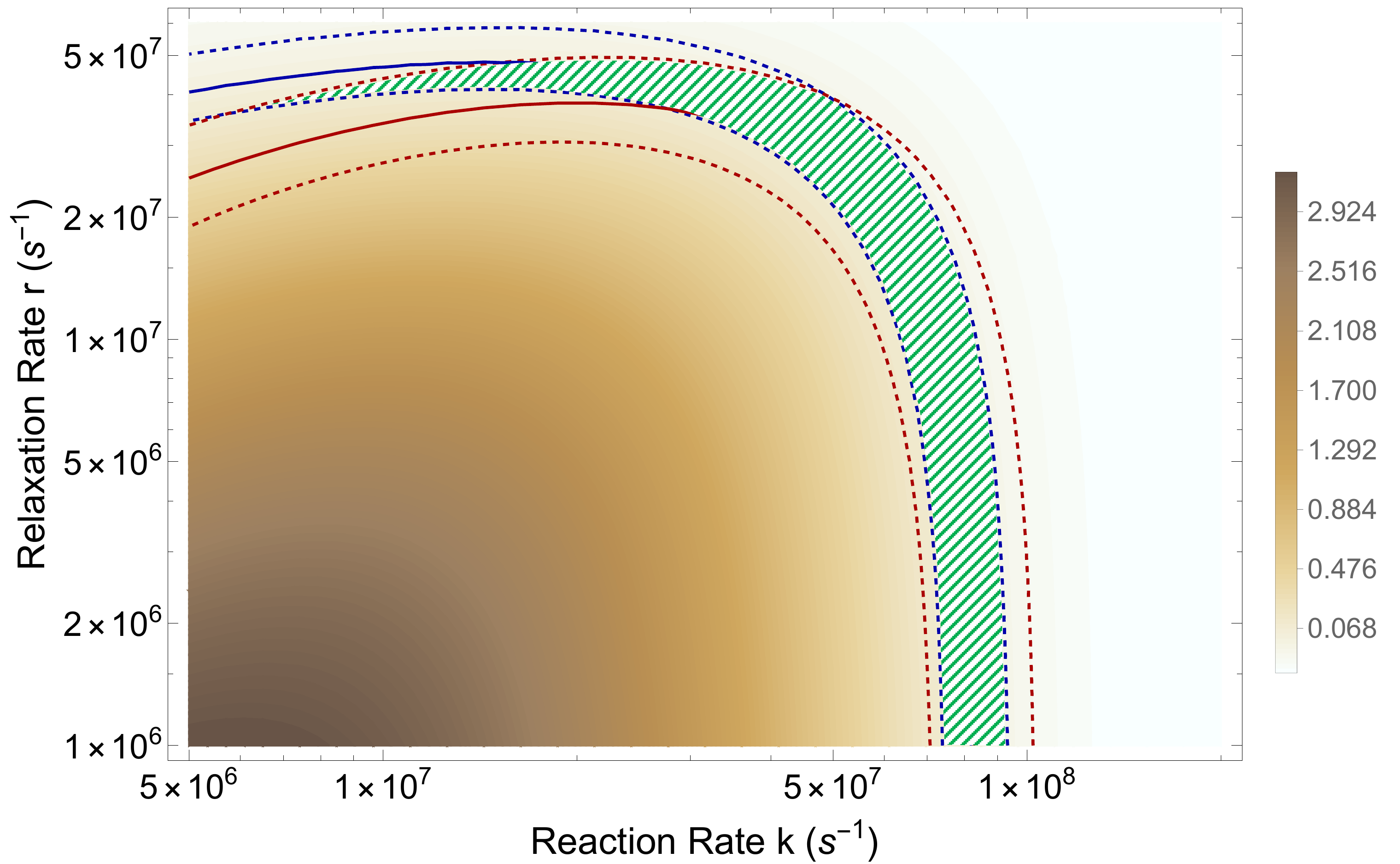}
    \caption{The RPM model can reproduce both magnetic field and lithium effects. The comparison between period changes due to applied magnetic fields  measured in the experiment \cite{yoshii2009cryptochrome}, $\Delta_{expt-MF}$, and obtained by the RPM model, $\Delta_{RPM-MF}$, where $\Delta_{MF}$ is the difference between period changes at 300 $\mu$T and 0 $\mu$T, $\Delta_{MF}=period_{300 \mu T}-period_{ 0 \mu T}$. The solid blue line indicates $\Delta_{expt-MF}-\Delta_{RPM-MF}=0$ h and the dashed blue line indicates the region where $|\Delta_{expt-MF}-\Delta_{RPM-MF}|\leq std_{expt-MF}=0.089$ h. The difference between period changes due to the lithium administration measured in the experiment \cite{dokucu2005lithium}, $\Delta_{expt-Li}$, and obtained by the RPM model, $\Delta_{RPM-Li}$ is presented by red lines. The solid red line indicates $\Delta_{expt-Li}-\Delta_{RPM-Li}=0$ h  and the dashed red line indicates the region where $|\Delta_{expt-Li}-\Delta_{RPM-Li}|\leq std_{expt-Li}=0.214$ h, $\Delta_{Li}=period_{30 mM}-period_{ 0  mM}$. The green shaded color indicates the regions where the RPM model
can reproduce both magnetic field \cite{yoshii2009cryptochrome} and lithium \cite{dokucu2005lithium} effects on \textit{Drosophila}'s CC. The parameters for calculating the period of the circadian clock are taken from Table  \ref{tab:tyson}, except that for the MF effects under blue light illumination $k_{p3}=0.085$ h$^{-1}$.}%
    \label{fig:cont}%
\end{figure}

\begin{figure}
     \begin{subfigure}{0.45\linewidth}
        \centering
        \includegraphics[width=1.\textwidth]{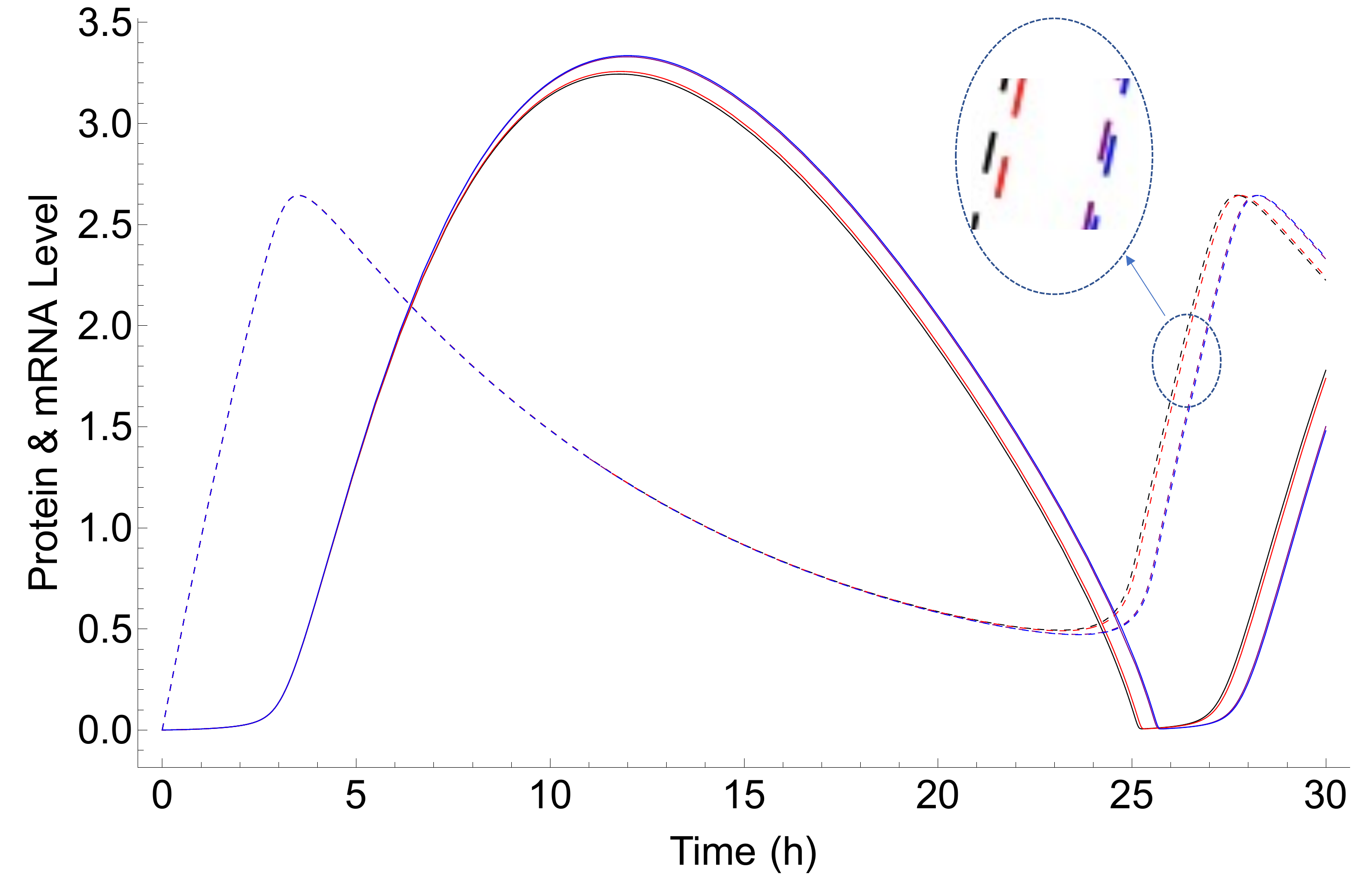}
        \caption{}
        \label{fig:pro/mrna-t-li}
    \end{subfigure}
    \hfill
\begin{minipage}{0.45\linewidth}
    \begin{subfigure}{\linewidth}
\includegraphics[width=1.\textwidth]{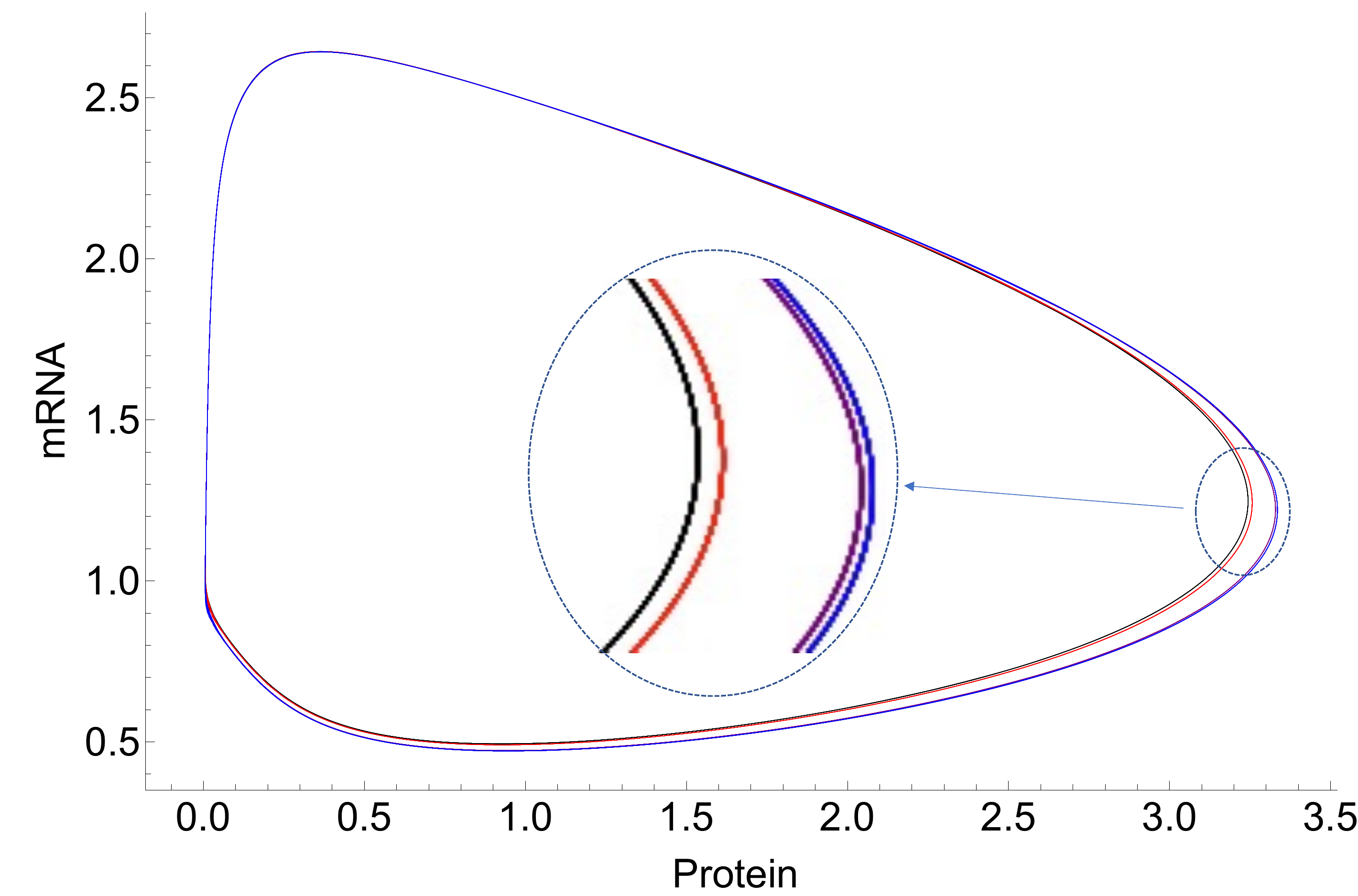}
        \caption{}
        \label{fig:pro-mrna-li}
    \end{subfigure}
\end{minipage}%
\caption{Lithium effects on the circadian clock are reproduced by the RPM model. a) Explicit time-dependence of the concentrations of protein [the solid lines] and mRNA [the dashed lines] and b) Parametric representations of oscillations in the concentrations of protein and mRNA, in \textit{Drosophila}'s circadian clock model using the parameters from Table \ref{tab:tyson}. The black, red, blue and purple colors indicate zero-lithium, \ce{^{6}Li}, \ce{^{7}Li}, and Li, respectively. Lithium administration prolongs the period of the clock, such that \ce{^{7}Li} has more potency than \ce{^{6}Li}.}
\label{fig:cc-li}
\end{figure}

\begin{figure}
     \begin{subfigure}{0.45\linewidth}
        \centering
        \includegraphics[width=1.\textwidth]{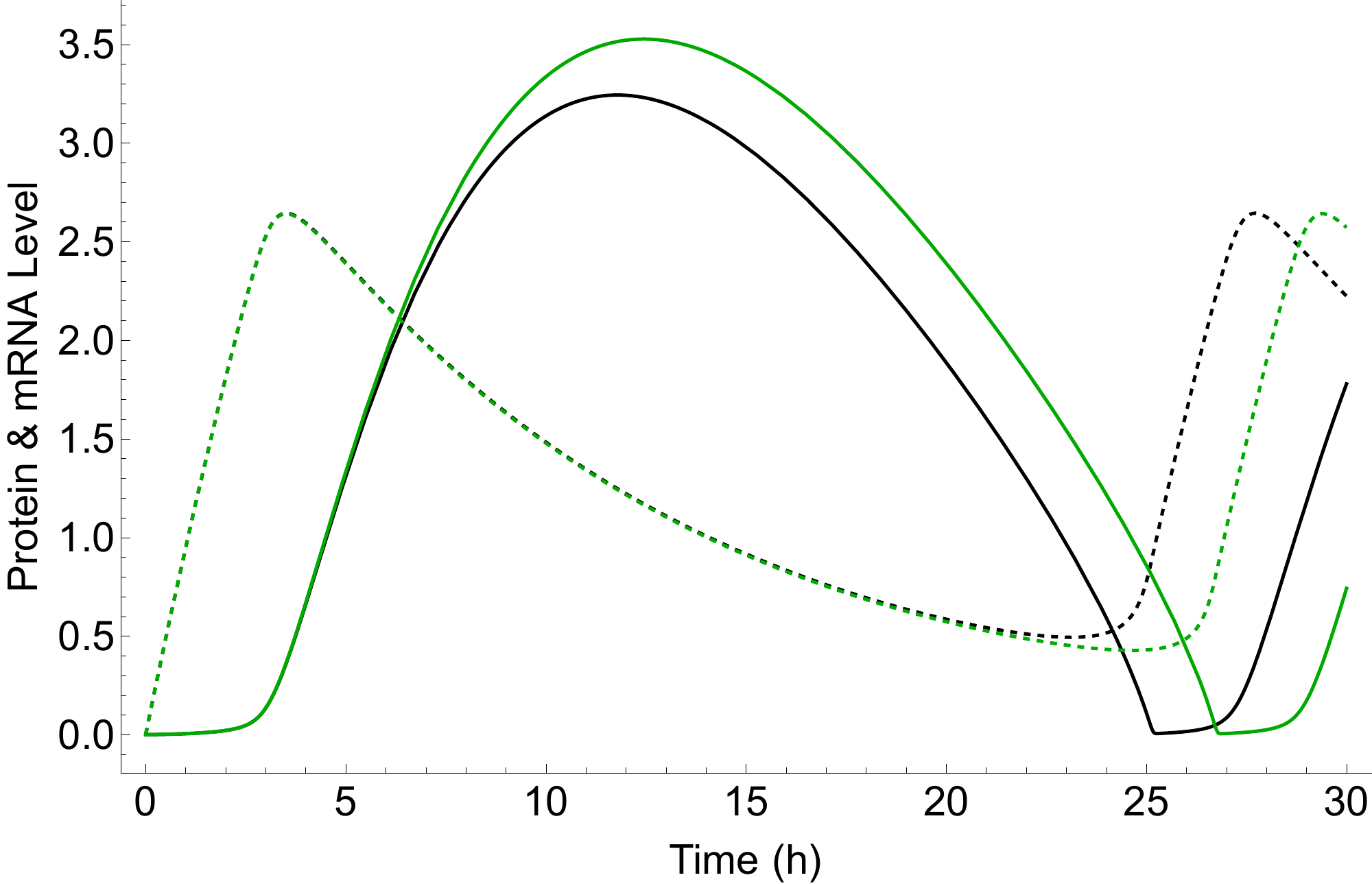}
        \caption{}
        \label{fig:pro/mrna-t-mf}
    \end{subfigure}
    \hfill
\begin{minipage}{0.45\linewidth}
    \begin{subfigure}{\linewidth}
\includegraphics[width=1.\textwidth]{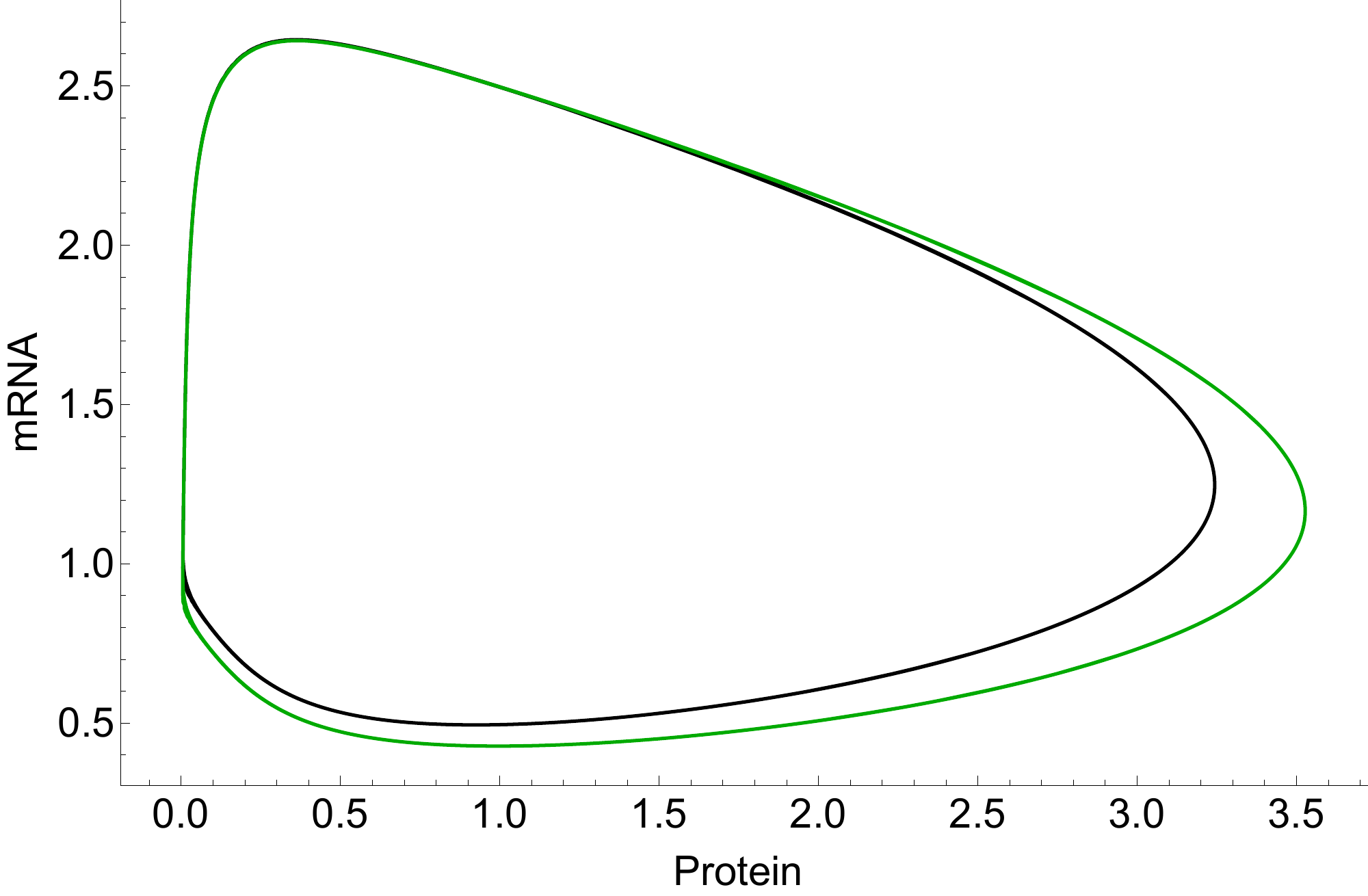}
        \caption{}
        \label{fig:pro-mrna-mf}
    \end{subfigure}
\end{minipage}%
\caption{Magnetic field effects on the circadian clock are reproduced by the RPM model. a) Explicit time-dependence of the concentrations of protein [the solid lines] and mRNA [the dashed lines] and b) Parametric representations of oscillations in the concentrations of protein and mRNA, in \textit{Drosophila}'s circadian clock model using the parameters from Table \ref{tab:tyson}, except $k_{p3}=0.085$ h$^{-1}$. The black and green colors indicate zero-MF and 300 $\mu$T MF effects, respectively.}
\label{fig:cc-mf}
\end{figure}

Based on the above considerations, here, we calculate the explicit effects of an applied magnetic field and the hyperfine interactions on the period of the CC. Using Eqs. \ref{eq:m}, \ref{eq:pt}, and \ref{eq:yeild-chem3}, we explored the parameter space of relaxation rate $r$ and recombination rate $k$ in order to find allowed regions for which our model can reproduce both experimental findings of static MF of 300 $\mu$T \cite{yoshii2009cryptochrome} and 30 mM of lithium \cite{dokucu2005lithium} effects on \textit{Drosophila}'s CC, which respectively lengthen the clock's period by 0.224$\pm$0.068 h and 0.567$\pm$0.11 h. The results are shown in Fig. \ref{fig:cont}. We find an allowed region where the model reproduces both experiments, see Fig. \ref{fig:cont}. The parameters for calculating the period of the circadian clock are taken from Table \ref{tab:tyson}. As discussed above, $k_{p3}$ corresponds to the degradation of TIM due to blue light exposure. For the MF effects under blue light illumination, we set  $k_{p3}=0.085$ h$^{-1}$ to obtain the control period of the circadian clock 25.8$\pm$0.14 h under blue light illumination observed in Ref. \cite{yoshii2009cryptochrome}. Fig. \ref{fig:cc-li} shows the effects of lithium on the rhythmicity of CC, such that \ce{^7{Li}} lengthens the period of the clock longer than \ce{^6{Li}}. For the effects of lithium on the circadian clock, the geomagnetic field of 50 $\mu$T is taken into account. Fig. \ref{fig:cc-mf} shows the effects of 300 $\mu$T MF on the CC.  

\begin{figure}
     \begin{subfigure}{0.47\linewidth}
        \centering
        \includegraphics[width=1.\textwidth]{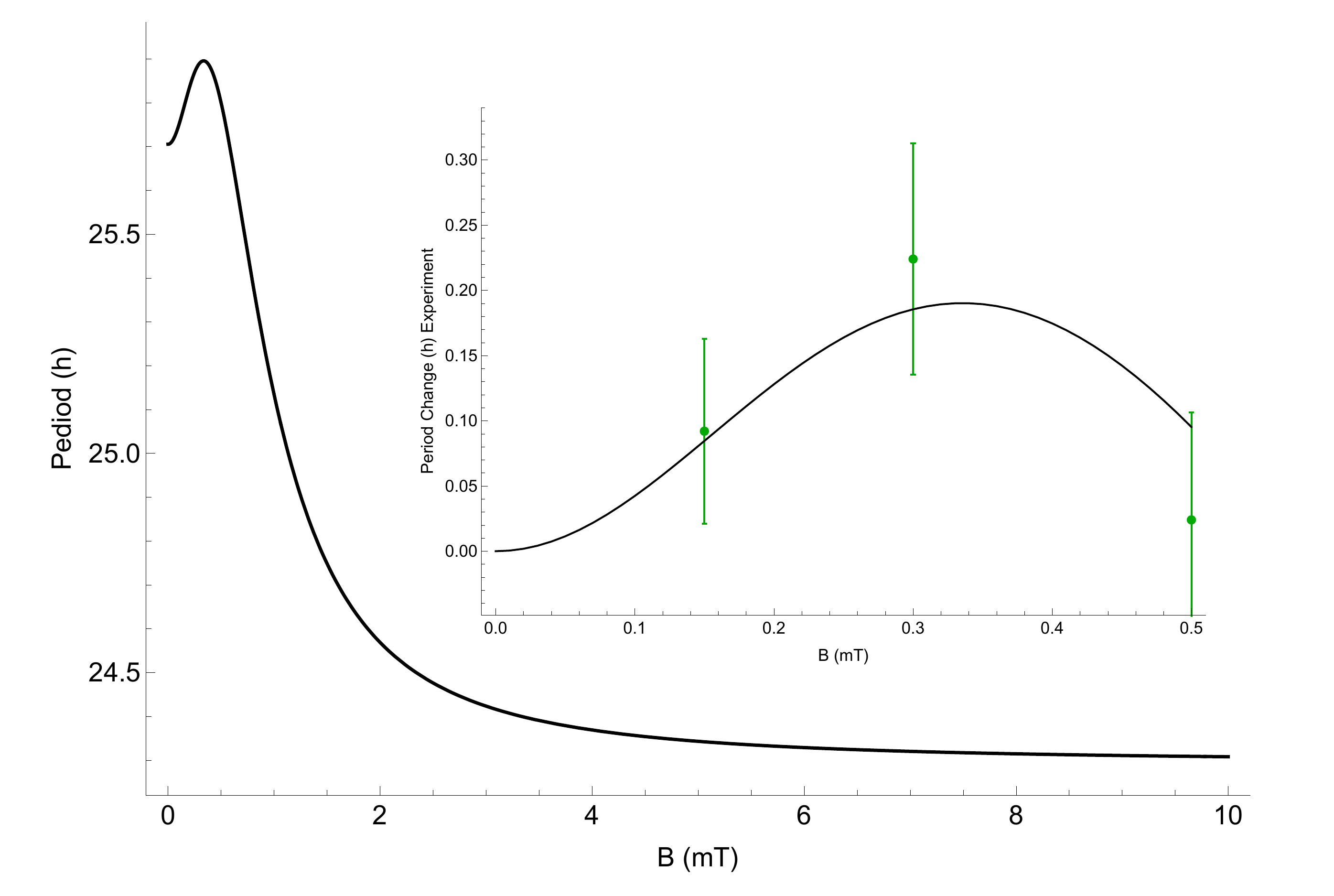}
        \caption{}
        \label{fig:fitting-mf}
    \end{subfigure}
    \hfill
\begin{minipage}{0.47\linewidth}
    \begin{subfigure}{\linewidth}
\includegraphics[width=01.\textwidth]{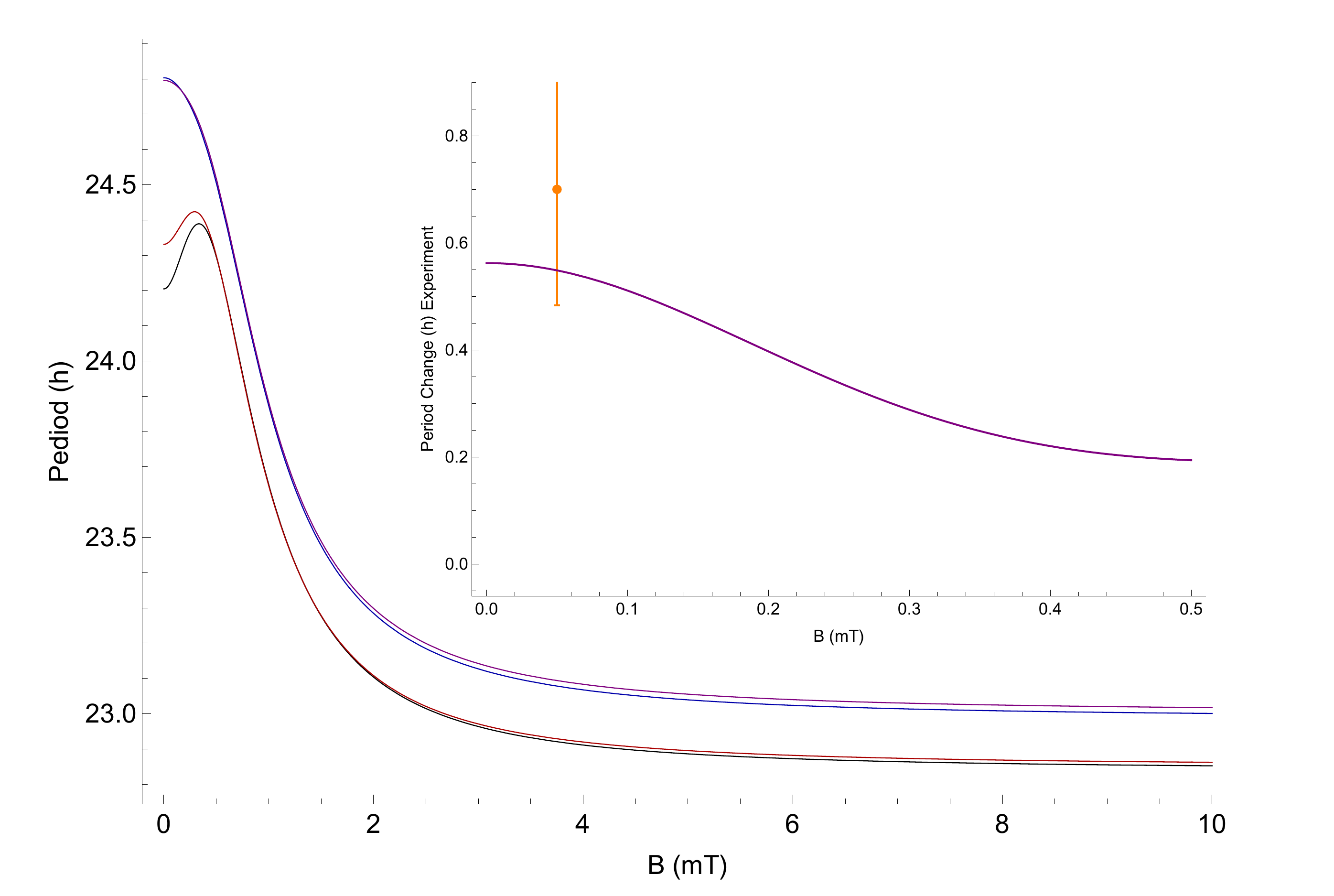}
        \caption{}
        \label{fig:fitting-li}
    \end{subfigure}
\end{minipage}%
\caption{The dependence of the period of \textit{Drosophila}'s circadian clock calculated by the RPM model on the static magnetic field strength $B$ without (a) and with (b) lithium effects for $a_A = 431.3$ $\mu$T, $a_B = a_{\ce{^7{Li}}} =-224.4$ $\mu$T, relaxation rate $r = 3\times10^7$ s$^{-1}$, and reaction rate $k= 4\times10^7$ s$^{-1}$. Higher magnetic field intensities shorten the period of the circadian clock. For the case without lithium effects (a), the applied magnetic field lengthens the period of the clock to a maximum in [280-360] $\mu$T and reduces effects at both lower and higher field strengths. The comparison between the dependence of the period on applied magnetic field  calculated by the RPM model [black line in the inset of plot (a)] and the experimental findings [green dots with error-bars] of Ref. \cite{yoshii2009cryptochrome}. b) The effects of \ce{{Li}} [purple], \ce{^6{Li}} [red], \ce{^7{Li}} [blue], and zero \ce{{Li}} [black]. The inset indicates the comparison between the effects of Li on the period of the  clock calculated by the RPM model [purple line] and the experimental findings [orange dots with error-bars] of Ref. \cite{dokucu2005lithium}. The results from the RPM fits into the uncertainty of the experimental data, such that $|\Delta_{expt}-\Delta_{RPM}|\leq std_{expt}$.}
\label{fig:fitting}
\end{figure}

The model here reproduces the dependence of the CC's period on the applied MF's strength and Li administration, shown in Fig. \ref{fig:fitting}. The model predicts that further increases in the intensity of the MF would shorten the period of the clock significantly. For the cases considering MF effects solely, for both the experimental data and the RPM model, the period reaches a maximum between 0 $\mu$T and 500 $\mu$T and exhibits reduced effects at both lower and slightly higher field strengths, shown in Fig. \ref{fig:fitting-mf}. For the cases of \ce{^6{Li}} or without lithium intake, the largest prolongation of the period occurs in the same range of magnetic field as well, shown in Fig. \ref{fig:fitting-li}. Another prediction of the model is that \ce{^7{Li}} prolongs the clock's period stronger than \ce{^6{Li}}, which has a smaller spin compared to \ce{^7{Li}}, see Fig. \ref{fig:fitting-li}. In this model, in the cases where lithium effects are considered, the geomagnetic effects of 50 $\mu$T are also considered. For the comparison between our model and the experimental data on the lithium effects, we assume that natural lithium was administered in the experiment \cite{dokucu2005lithium}. Fig. \ref{fig:fitting} shows that the dependence of the period on applied MFs and lithium effects calculated by the RPM model used in the present work is consistent with the experimental observations. We compare the maximum lengthening of the period in both the RPM model and experimental data \cite{yoshii2009cryptochrome}, $\Delta_{RPM}=0.154$ h and $\Delta_{expt}=0.22\pm 0.089$ h, respectively, where $\Delta=period_{max}-period_{min}$. The results from the RPM fall into the uncertainty of the experimental data, $|\Delta_{expt}-\Delta_{RPM}|\leq std_{expt}$.

\section*{Discussion}
In this project, we aimed to probe whether a RP model can explain the experimental findings for both the effects of static magnetic field \cite{yoshii2009cryptochrome} and lithium  \cite{dokucu2005lithium} on the circadian clock in \textit{Drosophila}. We showed how the quantum effects affect the rates, which then yields a change in the period of the clock. This is a significant step forward compared to the previous studies on xenon anesthesia \cite{Smith2021} and the lithium effects on hyperactivity \cite{zadeh2021entangled}, where the quantum effects were correlated to experimental findings without explicitly modeling the related chemical reaction networks. With a set of reasonable parameters, our model reproduces the experimental findings, as shown in Figs. \ref{fig:cont}, and \ref{fig:fitting}. In addition, this strengthens the previously proposed explanation for the effects of lithium on hyperactivity \cite{zadeh2021entangled} via the circadian clock.\par

We proposed that applied magnetic fields and nuclear spins of lithium influence the spin state of the naturally occurring radical pairs in the [\ch{FADH^{.}} ... \ch{O2^{.-}}] in the circadian clock. This is inspired by the observations that the \textit{Drosophila} circadian clock is altered by external magnetic fields \cite{yoshii2009cryptochrome,xue2021biological}, which is accompanied by modulations in the ROS level \cite{edgar2012peroxiredoxins,ehnert2017extremely}, and by lithium administration\cite{dokucu2005lithium}. Let us note that it has also been suggested that lithium exerts its effects by inhibiting Glycogen synthase kinase-3 (GSK-3) \cite{ryves2001lithium,padiath2004glycogen}, however, here the presence of RPs is the natural explanation for magnetic field effects, but their existence in GSK-3 requires experimental support.\par

Of note, there is a large body of evidence that ROS are involved in the context of magnetosensing and the circadian clock modulations \cite{sherrard2018low,sherrard2018low,sheppard2017millitesla,emery1998cry,arthaut2017blue,edgar2012peroxiredoxins,pei2019diurnal}. it has been shown that oscillating magnetic fields at Zeeman resonance can influence the biological production of ROS \textit{in vivo}, indicating coherent S-T mixing in the ROS formation \cite{Usselman2016}. Additionally, it has been observed that extremely low frequency pulsed electromagnetic fields cause defense mechanisms in human osteoblasts via induction of \ch{O2^{.-}} and \ch{H2O2} \cite{ehnert2017extremely}. Sherrard et al. \cite{sherrard2018low} observed that weak pulsed electromagnetic fields (EMFs) stimulate the rapid accumulation of ROS, where the presence of CRY was required \cite{sherrard2018low}. The authors of that work concluded that modulation of intracellular ROS via CR represents a general response to weak EMFs. Further, Sheppard et al. \cite{sheppard2017millitesla} demonstrated that MFs of a few millitesla can indeed influence transfer reactions in \textit{Drosophila} CRY. It has also been shown that illumination of \textit{Drosophila} CRY results in the enzymatic conversion of molecular oxygen to transient formation of superoxide \ch{O2^{.-}} and accumulation of hydrogen peroxide \ch{H2O2} in the nucleus of insect cell cultures \cite{emery1998cry}. These findings indicate the light-driven electron transfer to the flavin in CRY signaling \cite{arthaut2017blue}. \par

The feasibility for the \ch{O2^{.-}} radical to be involved in the RPM is a matter of debate in this scenario due to its likely fast spin relaxation rate $r$. Because of fast molecular rotation, the spin relaxation lifetime of \ch{O2^{.-}} is thought to be on the orders of $1$ ns \cite{Player2019,Hogben2009}. Nonetheless, it has also been pointed out that this fast spin relaxation can be decreased on account of its biological environment. Additionally, Kattnig et al.  \cite{Kattnig2017,Kattnig2017b} proposed that scavenger species around \ch{O2^{.-}} can also reduce its fast spin relaxation. Moreover, in such a model, the effects of exchange and dipolar interactions can also be minimized. \par

It is often assumed that in the RP complexes involving superoxide are formed in triplet states, as opposed to the case considered here. This is because the ground state of the oxygen molecule is a triplet state. The initial state for RP formation could also be its excited singlet state, which is also which is its excited state (and is also a biologically relevant ROS) \cite{kerver1997situ,miyamoto2014singlet,kanofsky1989singlet}. Further, the transition of the initial RP state from triplet to singlet could also take place due to spin-orbit coupling 
\cite{goushi2012organic,fay2019radical}. Let it be
also noted that this model could be adapted for other RP complexes in the CC, namely [\ch{FAD^{.-}} ... \ch{TrpH^{+.}}].\par

Our model predicts that increasing the intensity of the applied magnetic field will shorten the period of the clock. This is a significant new prediction of our model that would be very interesting to check. The isotopic-dependence of the period is another prediction of our present model, such that \ce{^7{Li}} lengthens the period of the clock longer than \ce{^6{Li}}. Experiments on mammals would also be of interest \cite{takahashi2017transcriptional,becker2004modeling,geier2005entrainment,hiwaki1998influence,kassahun2020perturbing}. \par

The circadian clock not only controls the rhythms of the biological processes, but it also has intimate connections to other vital processes in the body \cite{ruan2021circadian} and particularly in the brain \cite{schmidt2007time}. It has been suggested that environmental perturbations in the circadian period could increase the risk of selected cancers and hence the circadian clock could be a therapeutic target for cancer risks \cite{kelleher2014circadian}. It also appears that the way drugs function depends on the circadian clock \cite{nahmias2021circadian,Raz2005}. Notably, it has been shown that the circadian clock is vital for maintaining the anti-oxidative defense \cite{krishnan2008circadian}. Moreover, it has been suggested that the circadian clock could be a new potential target for anti-aging \cite{ulgherait2020circadian,acosta2021importance} and neurodegenerative disorders therapeutics \cite{mattis2016circadian}. Thus this project also paves a potential path to study other functionalities of the body and the brain connected to the circadian clock in the light of the RPM. \par

To sum up, our results suggest that quantum effects may underlie the magnetic field and lithium effects on the circadian clock. A similar mechanism is likely to be at the heart of magnetoreception in animals \cite{Mouritsen2018}, xenon-induced anesthesia \cite{Smith2021}, and lithium treatment for mania \cite{zadeh2021entangled}. Our work is thus another piece of evidence that quantum entanglement may play essential roles in the brain's functionalities \cite{Hameroff2014b,Fisher2015,Simon2019,Adams2020,Kumar2016,gauger2011sustained,bandyopadhyay2012quantum,cai2010quantum,kominis2012magnetic,pauls2013quantum,tiersch2014approaches,zhang2014sensitivity}.

\section*{Methods}
\subsection{Quantum effects and chemical oscillator}
The effects of applied magnetic fields and hyperfine interactions can be introduced, following the work of Player et al. \cite{player2021amplification}, by assuming that the FAD signaling of CRY to TIM and hence the TIM degradation in the CC process proceed by RPs, shown in Eq. \ref{eq:chem}:

\begin{ceqn}
\begin{equation}\label{eq:chem}
\ch{FAD^* <>[ $k_{\mathrm{1}}$ ][ $k_{\mathrm{-1}}$ ] RP ->[ $k_{\mathrm{f}}$ ] FADH^{.}},
\end{equation}
\end{ceqn}
where rate constants $k_1$ and $k_{-1}$ are assumed to conserve electron spin. External magnetic fields and the hyperfine interactions can influence the overall rate of production of \ch{FADH^{.}} by altering the extent and timing of coherent singlet/triplet interconversion in RP and so changing the probability that it reacts to give \ch{FADH^{.}} rather than returning to \ch{FAD^*}. Based on a spin dynamics calculation, one can describe the effect of applied magnetic fields and the HFIs on the kinetics of the CC simply by modifying the rate constant $k_{p3}$ \cite{hunter1996regulation,lee1996resetting,myers1996light,zeng1996light} which corresponds to the degradation of TIM in Eqs. \ref{eq:m} and \ref{eq:pt}. The spin dynamics of the RP in Eq. \ref{eq:chem} can be written as follows: 

\begin{ceqn}
\begin{equation}\label{eq:master}
\frac{d\hat{\rho}(t)}{dt}=-\hat{L} \hat{\rho}(t)+k_1 \frac{\hat{P}^S}{M}=-i[\hat{H},\hat{\rho}(t)]-\frac{k_{-1}}{2}\{\hat{P}^S,\hat{\rho}(t)\}-k_f \hat{\rho}(t)+k_1 \frac{\hat{P}^S}{M},
\end{equation}
\end{ceqn}
where $\hat{L}$ is the Liouvillian, $[...,...]$ and $\{...,...\}$ are the commutator and anti-commutator operators, $\hat{\rho}(t)$ is the spin density operator of the RP system; its trace, $Tr[\hat{\rho}(t)]$, equals the concentration of RPs divided by the fixed concentration of FAD$^*$ in Eq. \ref{eq:chem}. As RPs are short-lived intermediates, their concentrations are very low, and hence one can obtain the steady-state solutions as follows (see Ref. \cite{player2021amplification}): 

\begin{ceqn}
\begin{equation}\label{eq:yeild-chem1}
\Phi=\frac{k_f}{M}Tr[\hat{L}^{-1}\hat{P}^S],
\end{equation}
\end{ceqn}
where $\Phi$ is the singlet yield of the RPM. It is, therefore, possible to introduce the singlet yield of the RPM to the chemical reaction by modifying the rate $k_f$. In the CC Eqs. \ref{eq:m} and \ref{eq:pt} the corresponding rate is $k_{p3}$, which is 0.1 h$^{-1}$ for the natural cycle of the clock without blue light illumination and 0.085 h$^{-1}$ for blue light illumination. The singlet yield effects on $k_{p3}$ can be written as follows:

\begin{ceqn}
\begin{equation}\label{eq:yeild-chem4}
k'_{p3} \propto k_{p3} \frac{\Phi'_S}{\Phi_S},
\end{equation}
\end{ceqn}
where $k'_{p3}$, $\Phi_S$, and $\Phi'_S$ are the modified rate constant $k_{p3}$, the singlet yield with no quantum effects, and the singlet yield resulted from quantum effects due to the Zeeman and/or hyperfine interactions, respectively.

\section*{Data Availability}
The generated datasets and computational analysis are available from the corresponding author on reasonable request.

\bibliography{sample}

\section*{Acknowledgements}
The authors would like to thank Rishabh, Dennis Salahub, Wilten Nicola, Gabriel Bertolesi, Sarah McFarlane, and Nilakshi Debnath for their valuable input. The authors also acknowledge Compute Canada for its computing resources. This work was supported by the Natural Sciences and Engineering Research Council of Canada.

\section*{Author contributions statement}
H.ZH. and C.S. conceived the project; H.ZH. performed the calculations; H.ZH. and C.S. wrote the paper; C.S. supervised the project.

\section*{Competing Interests}
The authors declare no competing interests.

\end{document}